\documentclass{aa}

\usepackage{graphicx}
\usepackage{txfonts}
\usepackage{natbib}

\begin{document}

\title{The Orbit of GG Tau A%
\thanks{Based on observations collected at the European Southern
  Observatory, Chile, proposals number 072.C-0022, 078.C-0386, and
  384.C-0870}}

\author{R. K\"ohler}

\institute{%
	Max-Planck-Institut f\"ur Astronomie, K\"onigstuhl 17,
	69117 Heidelberg, Germany,
        \email{koehler@mpia.de}
\and
	Landessternwarte, Zentrum f\"ur Astronomie der Universit\"at Heidelberg,
	K\"onigstuhl, 69117 Heidelberg, Germany,
        \email{r.koehler@lsw.uni-heidelberg.de}
}

\date{Received 15 December 2010; accepted 12 April 2011}

\abstract{}{We present a study of the orbit of the pre-main-sequence
  binary system GG~Tau~A and its relation to its circumbinary disk,
  in order to find an explanation for the sharp inner edge of the
  disk.}
{Three new relative astrometric positions of the binary were obtained
  with NACO at the VLT.  We combine them with data from the
  literature and fit orbit models to the dataset.}
{We find that an orbit coplanar with the disk and compatible with the
  astrometric data is too small to explain the inner gap of the disk.
  On the other hand, orbits large enough to cause the gap are tilted
  with respect to the disk.
 If the disk gap is indeed caused by the stellar companion, then the
 most likely explanation is a combination of underestimated
 astrometric errors and a misalignment between the planes of the disk
 and the orbit.
}{}


\maketitle


\section{Introduction}

\object{GG~Tau} is a young quadruple system consisting of two binaries
\citep{leinert93}.  GG~Tau~A is a pair of low-mass stars separated by
about $0.25''$.  GG~Tau~B, located $10.1''$ to the south, is wider
($1.48''$) and less massive.
A circumbinary disk around GG~Tau~A has been extensively studied.
It was spatially resolved in both near infrared and millimeter
wavelength domains.  A detailed analysis of the velocity maps of the
disk found that it is in Keplerian rotation and constrained the
central mass to $1.28\pm0.07\,M_\odot$ \citep{guilloteau99}.

So far, orbital motion has not been detected in the GG~Tau~B binary
because of its long period.  However, the relative motion of the
components of GG~Tau~A has been observed for several years and has
already resulted in several orbit determinations \citep{mccabe2002,
tamazian2002, beust2005}. Because only a limited section of the orbit
has been observed, the authors generally have assumed that the orbit
is coplanar with the circumstellar disk ($i=37^\circ\pm1^\circ$).  The
resulting orbital parameters were all quite similar to each other,
with a semi-major axis of about 35\,AU.

The presence of the binary would be an obvious explanation for the
rather sharp inner edge of the disk located at 180\,AU.  The ratio of
the inner radius of the disk and the semi-major axis is about five.
However, \cite{artymowicz94} studied the effect of binary
systems on their circumbinary disks and found that this ratio should
range from about 1.7 (for circular orbits) to about 3.3 (for highly
eccentric binaries, $e=0.75$).  \cite{beust2005, beust2006} carried
out a similar study specifically for GG~Tau~A and came to the same
conclusion.  The binary orbit cannot explain the gap in the
circumbinary disk, unless its semi-major axis is about twice as large
as indicated by the astrometric data available.

In this paper, we present new relative astrometric measurements of
GG~Tau~A and derive estimates for its orbital parameters, with and
without the assumption that binary orbit and circumbinary disk are
coplanar.


\section{Observations and data reduction}

Astrometric measurements of GG~Tau~A have been published by several
authors \citep{duchene2004, ghez95, ghez97, hartigan2003, krist2002,
  leinert93, mccabe2002, roddier96, tamazian2002, white2001, woitas2001},
see \cite{beust2005} for an overview.
Here we report on new observations obtained with NAOS/CONICA (NACO for
short), the adaptive optics, near-infrared camera at the ESO Very
Large Telescope on Cerro Paranal, Chile \citep{rousset03, lenzen03}.
GG~Tau was observed on December 13, 2003 (PI: Leinert), November 20,
2006 (PI: Ratzka), and October 5, 2009 (PI: K\"ohler).
We use only imaging observations in the $K_s$ photometric band for the
orbit determination.
Integration times were 85\,sec per image in 2003, 24\,sec in 2006, and
60\,sec in 2009.  In 2003 and 2006, we took four images with the star
at different positions on the detector to facilitate creation of a
median sky image.  In 2009, 12 images were recorded at four
positions.

\begin{figure}[t]
\centerline{\includegraphics[width=0.8\hsize]{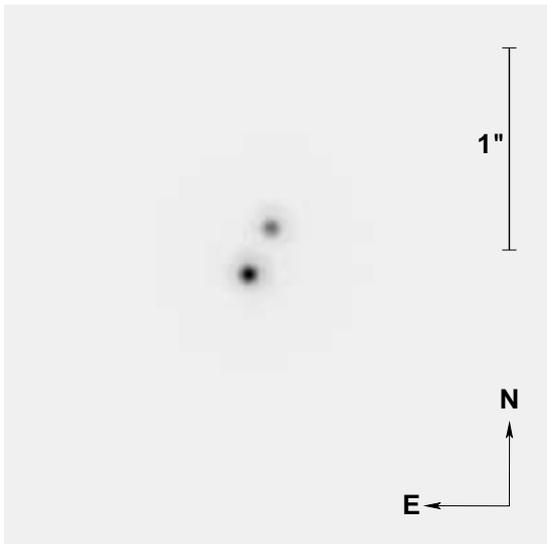}}
\caption{Image of GG Tau A obtained with NACO in October 2009.
  The separation between the two components is about 250\,mas.}
\label{NACOpic}
\end{figure}

The NACO images were sky subtracted with a median sky image, and
bad pixels were replaced by the median of the closest good neighbors.
Finally, the images were visually inspected for any artifacts or
residuals.  Figure~\ref{NACOpic} shows an example of the results.

The Starfinder program \citep{Diolaiti00} was used to measure the
positions of the stars.  The positions in several images taken during
one observation were averaged, and their standard deviation used to
estimate the errors.  To derive the exact pixel scale and orientation
of the detector, we took images of fields in the Orion Trapezium
during each observing campaign, and reduced them in the same way as
the images of GG~Tau.  The measured positions of the cluster stars
were compared with the coordinates given in \cite{mccaughrean94}.
The mean pixel scale and orientation were computed from a global fit
of all star positions.  The scatter of values derived from star pairs
was used to estimate the errors.  The errors of the calibration are
usually comparable to or larger than the errors of the measured
positions of the science target, indicating the importance of a proper
astrometric calibration.

The calibrated separations and position angles of GG~Tau~A appear in
Table~\ref{ObsTab}, together with the data taken from the literature.
If one or both components of GG~Tau~B were within the field-of-view,
then we also measured their positions.  The results appear in
Table~\ref{ObsBTab}.  The main conclusion is that there has been no
significant change in the relative position since the first
measurement published in \cite{leinert93}.


\begin{table}[t]
\setlength{\tabcolsep}{3pt}
\caption{Astrometric measurements of GG Tau Aa -- Ab}
\label{ObsTab}
\begin{center}
\begin{tabular}{ll@{${}\pm{}$}lr@{${}\pm{}$}ll}
\noalign{\vskip1pt\hrule\vskip1pt}
 Date (UT) 	& \multicolumn{2}{c}{$d$ [mas]}	& \multicolumn{2}{c}{PA~$[^\circ]$} & Reference\\
\noalign{\vskip1pt\hrule\vskip1pt}
 1990 Nov  2	& $255. $&$10.$ & $  9.\phantom{0}$&$2. $ & \cite{leinert93}\\	
 1991 Oct 21	& $260. $&$10.$ & $  2.\phantom{0}$&$1. $ & \cite{ghez95}\\	
 1993 Dec 26	& $260. $&$10.$ & $  3.\phantom{0}$&$2. $ & \cite{roddier96}\\	
 1994 Jan 27	& $246. $&$4. $ & $357.8$&$0.4$ & \cite{woitas2001}\\		
 1994 Jul 25	& $250.2$&$2.6$ & $358.8$&$0.45$& \cite{ghez97}	\\		
 1994 Sep 24	& $258. $&$4. $ & $357.\phantom{0}$&$2. $ & \cite{ghez95} \\	
 1994 Oct 18	& $242. $&$3. $ & $  0.9$&$0.5$ & \cite{ghez95} \\		
 1994 Dec 22	& $239. $&$5. $ & $357.2$&$2. $	& \cite{roddier96}\\	
 1995 Oct  8	& $247. $&$4. $ & $356.9$&$0.7$	& \cite{woitas2001}\\	
 1996 Sep 29	& $245. $&$4. $ & $355.5$&$0.4$	& \cite{woitas2001}\\	
 1996 Dec  6	& $243.6$&$4.6$ & $354.9$&$1.3$	& \cite{white2001}\\	
 1997 Sep 27	& $250. $&$3. $ & $354.3$&$1. $	& \cite{krist2002}\\	
 1997 Oct 10	& $248. $&$2. $ & $353.9$&$0.4$	& \cite{mccabe2002}\\	
 1997 Nov 16	& $247. $&$5. $ & $353.6$&$0.4$	& \cite{woitas2001}\\	
 1998 Oct 10	& $260. $&$4. $ & $350.7$&$0.4$	& \cite{woitas2001}\\	
 2001 Jan 21	& $248. $&$14.$ & $348.6$&$2.4$	& \cite{hartigan2003}\\	
 2001 Feb  9	& $245. $&$4. $ & $348.7$&$0.3$	& \cite{tamazian2002}\\	
 2002 Dec 12	& $250.7$&$1.5$ & $346.0$&$1.5$	& \cite{duchene2004}\\	
 2003 Dec 13	& $250.7$&$0.8$	& $344.2$&$0.1$ & this work\\
 2006 Nov 20	& $252.3$&$0.7$	& $339.0$&$0.1$ & this work\\
 2009 Oct  5	& $252.5$&$0.3$	& $334.5$&$0.1$ & this work\\
\noalign{\vskip1pt\hrule\vskip1pt}
\end{tabular}
\end{center}
\end{table}

\begin{table}[t]
\setlength{\tabcolsep}{3pt}
\caption{Astrometric measurements of GG Tau B}
\label{ObsBTab}
\begin{center}
\begin{tabular}{lcr@{${}\pm{}$}lr@{${}\pm{}$}l}
\noalign{\vskip1pt\hrule\vskip1pt}
 Date (UT) 	& Pair	& \multicolumn{2}{c}{$d$ [arcsec]} & \multicolumn{2}{c}{PA~$[^\circ]$}\\
\noalign{\vskip1pt\hrule\vskip1pt}
 2006 Nov 20	& Bb--Ba &  $1.460$&$0.002$	& $134.9$&$0.1$ \\
		& Aa--Ba & $10.07 $&$0.01$	& $185.4$&$0.1$ \\
 2009 Oct  5	& Aa--Ba & $10.09 $&$0.01$	& $185.5$&$0.1$ \\
\noalign{\vskip1pt\hrule\vskip1pt}
\end{tabular}
\end{center}
\end{table}


\section{Determination of orbital elements}

\citet{mccabe2002} and \citet{beust2005} have determined the orbital
elements of GG Tau Aa-Ab from the average position and velocity of the
companion.
Together with the system mass \citep{guilloteau99}, position and
velocity on the sky comprise five measurements.  Since orbital
elements are seven unknowns, their computation requires the additional
assumption that the orbit and the circumbinary disk are coplanar.

In this work, we employed a different approach.  We fit orbit models
to the observations and searched for the model with the minimum
$\chi^2$.  In the end, we wanted to use a Levenberg-Marquardt
algorithm \citep{press92}.  However, the results of this algorithm
depend strongly on the chosen start values.  To avoid any bias for a
particular orbit, we carried out a preliminary fit that consists of a
grid search in eccentricity $e$, period $P$, and time of periastron
$T_0$.  Singular value decomposition was used to solve for the
remaining four elements.  The result is a grid of $\chi^2$ as function
of $e$ and $P$.  Since we were interested in the semi-major axis $a$
of the orbit, this was converted onto a $a$-$e$-grid by finding the
orbit model with the closest $a$ for each grid point.  The grid spans
a range from 20 to 200\,AU in $a$, and from 0 to 0.99 in $e$.

To convert the measured separations into AU, a distance of 140\,pc was
adopted \citep{Elias78}.


\begin{figure}[tp]
  \includegraphics[angle=90,width=0.83\hsize]{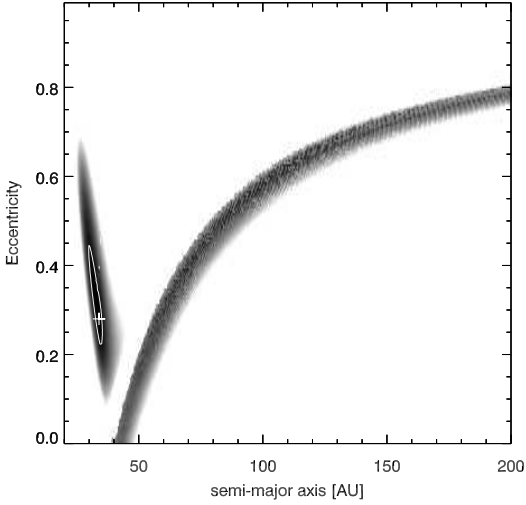}
\caption{$\chi^2$ as function of $a$ and $e$ for orbit models that are
  coplanar with the circumbinary disk (Sect.~\ref{coplanSect}).
  The cross at $a=34\rm\,AU$, $e=0.28$ marks the minimum, the contour
  line around it encircles the 99.7\,\% confidence region
  (corresponding to $3\sigma$ in the case of normally distributed
  errors).  The areas in various shades of gray are within the
  $5\sigma$ confidence region, i.e.\ orbit models in the white area
  can be excluded with $5\sigma$ confidence.
}
\label{ChiWithDiskFig}
\end{figure}

\begin{figure}[tp]
  \includegraphics[angle=90,width=0.83\hsize]{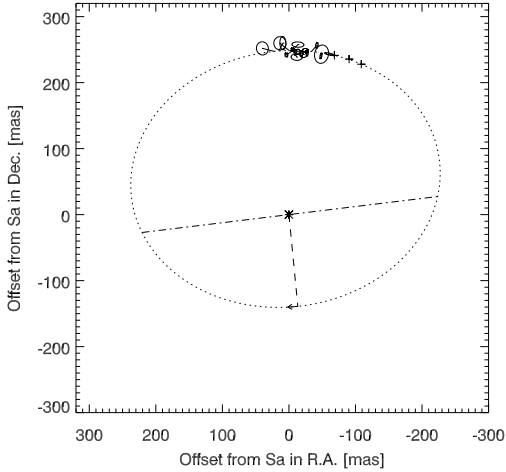}
\caption{Best-fitting orbit model if the orbit is constrained to be
  coplanar with the circumbinary disk (Sect.~\ref{coplanSect}).
  The observed positions are marked by their error ellipses and
  lines connecting the observed and calculated position at the
  time of the observations.  The new observations with NACO are marked
  by crosses. Their errors are too small to be discernible.
  The dash-dotted line indicates the line of nodes, the dashed line
  the periastron, and the arrow shows the direction of the orbital motion.
}
\label{OrbitWithDiskFig}
\end{figure}


\begin{figure}[tp]
  \includegraphics[angle=90,width=0.83\hsize]{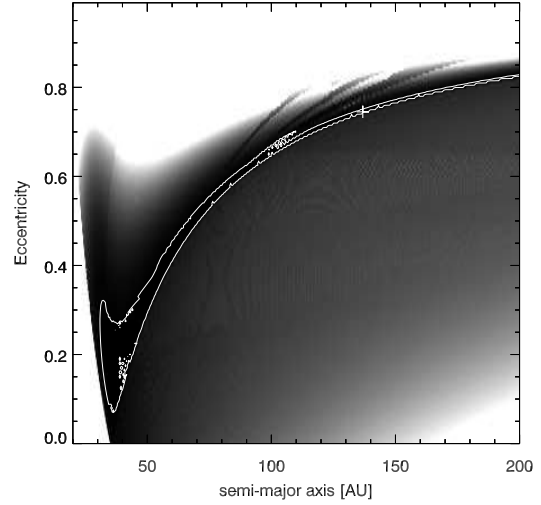}
\caption{$\chi^2$ as function of $a$ and $e$ for orbit models that are
  not necessarily coplanar with the circumbinary disk
  (Sect.~\ref{nocoplaSect}).  As in Fig.~\ref{ChiWithDiskFig}, the
  minimum $\chi^2$ is marked by the cross, but the contour line around
  it encircles the 68.3\,\% confidence region ($1\sigma$).  The areas
  in shades of gray are within the $5\sigma$ confidence region.  The
  jagged shape of the contour line is most likely caused by numerical
  effects.  }
\label{ChiWithoutDiskFig}
\end{figure}

\begin{figure}[tp]
  \includegraphics[angle=90,width=0.83\hsize]{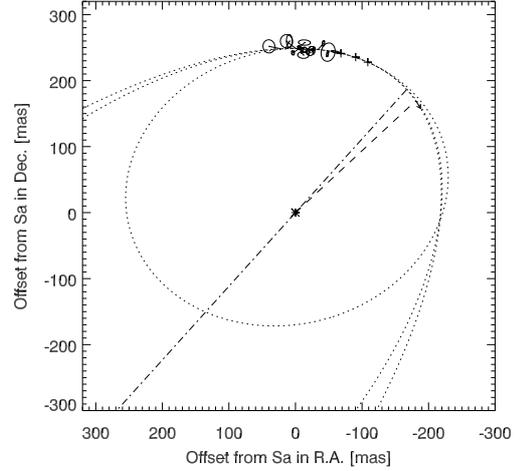}
\caption{Three exemplary orbit models that fit the astrometric data
  and the system mass, but are not coplanar with the disk.  The
  semi-major axes of the orbits are 35\,AU, 85\,AU, and 137\,AU.
  The last orbit has the minimal $\chi^2$, and its line of nodes is
  marked by the dash-dotted line and its periastron by the dashed
  line.  The observed positions are marked by their error ellipses and
  lines connecting the observed and calculated position at the
  time of the observations.  The new observations with NACO are marked
  by crosses. Their errors are too small to be discernible.
}
\label{OrbitWithoutDiskFig}
\end{figure}


\subsection{Orbits coplanar with the disk}
\label{coplanSect}

First, we searched for an orbit matching all the information available,
i.e.\ the astrometric position, the total mass, and the orientation of
the disk plane.  We assumed that disk and orbit are coplanar, orbits
without this constraint are discussed in the next section.

The $\chi^2$ that we try to minimize is
\begin{eqnarray}
\chi^2 &=&
  \sum_i \left(\vec r_{i,\rm obs} - \vec r_{i,\rm model}
			\over \Delta\vec r_{i,\rm obs}\right)^2
   + \left(M_{\rm est}-M_{\rm model} \over\Delta M_{\rm est}\right)^2
\nonumber\\
&& + \left(i_{\rm disk}-i_{\rm model}\over\Delta i_{\rm disk}\right)^2
   + \left(\Omega_{\rm disk}-\Omega_{\rm model}
  			\over\Delta\Omega_{\rm disk}\right)^2,
\label{ChiWithDiskEq}
\end{eqnarray}
where $\vec r_{i,\rm obs}$ and $\vec r_{i,\rm model}$ are the measured
and predicted position at the time of observation $i$, and $\Delta\vec
r_{i,\rm obs}$ is the error of the measurement.  Here, $M_{\rm est}$
is the measured system mass
\citep[$1.28\pm0.07\,M_\odot$,][]{guilloteau99}, and $M_{\rm model}$
the system mass predicted by the orbit model.
Then, $i_{\rm disk}$ and $\Omega_{\rm disk}$ are the inclination and
position angle (PA) of the ascending node of the orbit of a disk particle,
$\Delta i_{\rm disk}$ and $\Delta\Omega_{\rm disk}$ are their errors.
The inclination of the disk is $37\pm1^\circ$ \citep{guilloteau99}, but
it is in retrograde rotation, so $i_{\rm disk} = 180-37 = 143^\circ$.
The PA of the minor axis of the disk is $7\pm2^\circ$
\citep{guilloteau99}, therefore $\Omega_{\rm disk} = 277\pm2^\circ$
\citep[the ascending node is defined as the point in the orbit where the
object is receding from the observer most rapidly, e.g.][]{hilditch01}.

Equation\,\ref{ChiWithDiskEq} was minimized by a Levenberg-Marquardt
algorithm \citep{press92}.  The starting points for the algorithm were
taken from the preliminary fit described in the previous section.  We
kept $a$ and $e$ fixed to preserve the grid in these two variables.
The resulting $\chi^2$ distribution is depicted in
Fig.~\ref{ChiWithDiskFig}.  There is a clear minimum at $a=34\rm\,AU$
and $e=0.28$, while orbits with $a>36\rm\,AU$ can be excluded on the
$3\sigma$ level.  This is in perfect agreement with previous orbit
determinations.  The reduced $\chi^2$ at the minimum is 3.05, which
indicates a less-than-perfect fit.  Figure~\ref{OrbitWithDiskFig} shows
the orbit with the minimum $\chi^2$, together with the measurements of
the relative positions, and Table~\ref{OrbElTab} lists the orbital
elements.


\begin{table*}[t]
\caption{Parameters of the best orbital solutions.}
\label{OrbElTab}
\begin{center}
\begin{tabular}{lr@{}lr@{}lr@{}l}
\noalign{\vskip1pt\hrule\vskip1pt}
Orbital Element			& \multicolumn{2}{c}{Orbit coplanar} & \multicolumn{2}{c}{Orbit not}	& \multicolumn{2}{c}{most plausible orbit}\\
				& \multicolumn{2}{c}{with disk}  & \multicolumn{2}{c}{coplanar w.~disk} & \multicolumn{2}{c}{(see Sect.~\ref{DiscussSect})}\\
\noalign{\vskip1pt\hrule\vskip3pt}
Date of periastron $T_0$	& $2477680$ & $\,^{+690}_{-270}$  & $2460050$ & $\,^{+430}_{-500}$	& \quad$2463400$ & $\,^{+1470}_{-5420}$\\[3pt]
				& (July 2071)\span		 & (April 2023)\span			& (June 2032)\span		\\[2pt]
Period $P$ (years)		& $  162  $ & $\,^{+62}_{-15}$	 & $  1400$ & $\,^{+17700}_{-1300}$	& $ 403 $ & $\,^{+67}_{-32}$	\\[3pt]
Semi-major axis $a$ (mas)	& $  243  $ & $\,^{+38}_{-10}$	 & $   977$ & $\,^{+96}_{-90}$	  	& $ 429 $	\\[3pt]
Semi-major axis $a$ (AU)	& $   34  $ & $\,^{+5.9}_{-2.8}$  & $   137$ & $\,^{+17}_{-16}$	  	& $  60 $	\\[3pt]
Eccentricity $e$		& $   0.28$ & $\,^{+0.05}_{-0.14}$& $  0.75$ & $\,^{+0.03}_{-0.03}$	& $  0.44$ & $\,^{+0.02}_{-0.03}$\\[3pt]
Argument of periastron
	$\omega$ ($^\circ$)	& $  91   $ & $\,^{+4}_{-13}$	 & $     8$ & $\,^{+7}_{-9}$		& $  19  $ & $\,^{+9}_{-10}$	\\[3pt]
P.A. of ascending node
	$\Omega$ ($^\circ$)	& $ 277   $ & $\,^{+2.0}_{-2.0}$  & $   318$ & $\,^{+10}_{-7}$		& $ 131  $ & $\,^{+13}_{ -8}$	\\[3pt]
Inclination $i$ ($^\circ$)	& $ 143   $ & $\,^{+1.3}_{-1.0}$  & $   128$ & $\,^{+7}_{-4}$		& $ 132.5$ & $\,^{+1.0}_{-2.5}$	\\[3pt]
Angle between orbit and disk	& $   0.02$ & $\,\pm1.9$	 & $  31.8$ & $\,\pm1.5$		& $  24.9$ & $\,\pm1.7$	\\
\noalign{\hrule\vskip1pt}
\end{tabular}
\end{center}
\end{table*}


To test whether the astrometric errors were underestimated, we
repeated the procedure, but enlarged the errors of the observations by
a factor of 3.  This lowers $\chi^2$ in general, but does not result
in significant changes of the shape of the $\chi^2$-plane as function
of $a$ and $e$.  The best-fitting orbit has now $a=40\rm\,AU$ and
$e=0.13$.  Also, because of the lower $\chi^2$, many orbits with
$a>36\rm\,AU$ (up to the end of the grid at $a=200\rm\,AU$) are within
the 99.7\,\% confidence region (which corresponds to $3\sigma$ in the
case of a normal distribution).  However, it appears unlikely that the
authors of all astrometric data underestimated their errors by such a
large factor, and orbits large enough to cause the disk gap are still
only marginally consistent with the data.


\subsection{Orbits with no constraint on their orientation}
\label{nocoplaSect}

In this section, we remove the constraint that the orbit has to be in
the same plane as the circumbinary disk.  The only constraints are
therefore the astrometric measurements, and the total mass of the
binary.  Then, $\chi^2$ is given by (using the same symbols as in
Eq.~\ref{ChiWithDiskEq})
\begin{equation}
\chi^2 =
  \sum_i \left(\vec r_{i,\rm obs} - \vec r_{i,\rm model}
			\over \Delta\vec r_{i,\rm obs}\right)^2
   + \left(M_{\rm est}-M_{\rm model} \over\Delta M_{\rm est}\right)^2.
\label{ChiWithoutDiskEq}
\end{equation}

Figure~\ref{ChiWithoutDiskFig} shows the result of minimizing the
$\chi^2$ given by Eq.~\ref{ChiWithoutDiskEq}.  The formal minimum is
at $a = 137\rm\,AU$, $e=0.75$ (Table~\ref{OrbElTab}), with a reduced
$\chi^2$ of 3.2.  It is highly unlikely that the true orbit has such a
large semi-major axis and high eccentricity.
However, the minimum is very shallow, and no semi-major axis larger
than about 30\,AU can be excluded, not even at the $1\sigma$ level.
Figure~\ref{OrbitWithoutDiskFig} shows the orbit with the formally
minimal $\chi^2$, and two orbits that result in the best fit if the
semi-major axis is held fixed at 35\,AU and 85\,AU, respectively.  All
three orbits fit the measured data reasonably well, demonstrating that
the semi-major axis is not well constrained by the astrometric data.


\section{Discussion and conclusions}
\label{DiscussSect}

If we require the orbit model to lie in the same plane as the circumbinary
disk, then orbits consistent with the astrometric data are not large
enough to explain the gap in the disk.
On the other hand, if we consider orbits that are not coplanar with
the disk, then the astrometric data only provides a very weak
constraint for the semi-major axis.  This means that we can easily
find orbits that are consistent with the measured positions and with
the size of the gap in the circumbinary disk.

According to \cite{artymowicz94}, an orbit with eccentricity
$e\approx 0.4\ldots0.5$ can open a disk gap that is a factor of about
3 larger than its semi-major axis. For our disk with an inner edge at
about 180\,AU, a semi-major axis of 60\,AU would suffice.
Figure~\ref{OrbitWithoutDiskFig} shows that orbits with $a=60\rm\,AU$
should have an eccentricity of $0.4\ldots0.45$ to match the
astrometric data.  We consider this to be the most plausible orbit,
given the constraints from the astrometric data and the size of the
disk gap.  Its orbital elements appear in the rightmost column of
Table~\ref{OrbElTab}.

\cite{beust2005} have already discussed noncoplanar solutions for the
binary orbit.\footnote{They use a slightly different notation, where
  the position angle of the ascending node is replaced by the
  position angle of the projection of the rotation axis of the orbit
  onto the plane of the sky.  The difference between the two position
  angles is exactly $90^\circ$.}
Assuming $a = 62\rm\,AU$ and $e = 0.35$, they find
$i = 125.4^\circ$ and four solutions for $\Omega$: $114.3^\circ$,
$-65.7^\circ$, $74.5^\circ$, and $-105.6^\circ$. The first solution
differs from our most plausible orbit by $7^\circ$ in inclination and
$16.7^\circ$ in $\Omega$, which is a reasonable agreement given the
large uncertainties.

\begin{figure}[t]
  \includegraphics[angle=90,width=\hsize]{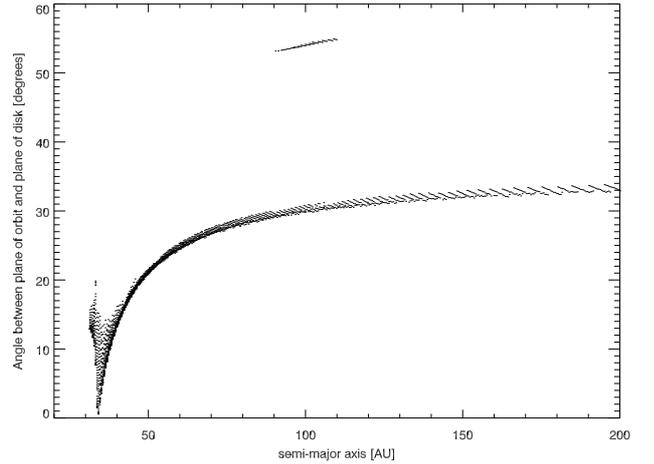}
  \caption{Angle between the plane of the binary orbit model and the
    plane of the circumbinary disk.  Plotted are the angles for all
    orbits within $1\sigma$ of the best fit, i.e.\ all the orbits
    within the white contour line in Fig.~\ref{ChiWithoutDiskFig}.
    These are orbit models that were not constrained by the
    orientation of the disk.}
  \label{MisalignFig}
\end{figure}

How large is the misalignment between the plane of the orbit and the
plane of the disk?  Figure~\ref{MisalignFig} shows the angle between
orbit and disk plane as a function of the semi-major axis of the orbit.
The relative astrometry of the two stars contains no information about
the sign of the inclination (whether the orbit is tilted towards or
away from the observer).  In the creation of Fig.~\ref{MisalignFig},
we adopted in each case the inclination that resulted in the smaller
angle between disk and orbit.


Most orbit models are tilted by less than about 35$^\circ$ with
respect to the disk.  Our most plausible orbit is inclined by
about 25$^\circ$, a significant misalignment.  \citet{beust2005}
point out that the disk should show a warped structure if it is not
coplanar with the binary orbit.  This has not been detected, making
such an orbit unlikely, although it cannot be ruled out.
\cite{beust2006} carried out some simulations of the dynamical
behavior of the disk for the noncoplanar orbits found by
\cite{beust2005} and a few possible orbits for the outer companion
GG~Tau~B. In all the simulations with the orbit similar to our most
plausible orbit (designated AA5 in \citealt{beust2006}), the disk
tends to assume an open-cone shape with an opening angle of
$\sim30^\circ$. This state is reached after 15 million years at the
end of simulations. These results suggest that the GG~Tau system
we observe today is only a transient feature.

GG~Tau is only about 1 million years old \citep{white2001}, so it is
possible that we happen to observe the disk just before it dissolves
into an open cone.  We can only speculate how the system got into this
unstable state.  It is well known that stars experience strong
gravitational interactions early in their lifetimes, even if they form
in small ensembles of only three to five stars
\citep[e.g.][]{SterzikDurisen98}.  These interactions can lead to
catastrophic changes in binary orbits and even to the ejection of
stars.  The four stars in the GG~Tau system would be enough to cause
such events, unless they are in a stable configuration.
Unfortunately, we have no kinematic information about the orbit of
GG~Tau~B, which is not surprising, since we expect an orbital period
on the order of 40000\,years (based on the projected separation of
1400\,AU).
It is conceivable that GG~Tau has recently suffered a gravitational
interaction and is currently in a transient, unstable state.
However, a gravitational interaction that changes the orbit of
GG~Tau~A should also have an effect on the circumbinary disk, making
it highly unlikely that the disk could maintain the planar structure
we see.

On the other hand, the orbital elements derived from the astrometric
data have rather large uncertainties.  For example, the $1\sigma$
confidence interval for the inclination of the orbit with
$a\approx85\rm\,AU$ ranges from $115^\circ$ to $158^\circ$ (based on
$\chi^2$ as function of inclination).  The errors of the angle between
orbit and disk should be comparable, although not identical, since the
angle between orbit and disk also depends on the orientation of the
line of nodes.

In summary, we do not have the final answer about the relation between
the orbit of GG~Tau~A and its circumbinary disk.  An orbit coplanar
with the disk could only cause the inner gap of the disk if the errors
of the astrometric measurements are much larger than estimated.
An orbit inclined to the plane of the disk would be compatible with
both the astrometric data and the disk gap, but it should cause
visible distortions in the disk structure.
An explanation for the fact that no distortions in the disk have been
detected could be that the orbit GG~Tau~A has only been changed
recently, although any effect that can change the orbit of the stars
should also disturb the structure of the disk.
On the other hand, we should not forget the possibility that the gap
in the disk is not related to GG~Tau~Ab, but some hitherto unknown
companion.  However, another companion would be pure speculation.

The most likely explanation seems to be a combination of slightly
underestimated astrometric errors and a (small) misalignment between
the planes of the orbit and the circumbinary disk.  More observations
over a larger section of the binary orbit are needed.


\begin{acknowledgements}
  I thank the referee Herve Beust for his comments and suggestions
  that helped to improve the paper.
\end{acknowledgements}

\bibliographystyle{bibtex/aa}
\bibliography{16327}

\end{document}